\documentclass[journal]{IEEEtran}
\usepackage{amsmath,amssymb,amsfonts}
\usepackage{amsthm}
\usepackage{mathrsfs}
\usepackage{algorithmic}
\usepackage{subfigure}
\usepackage{graphicx}
\usepackage{textcomp}
\usepackage{cite}
\usepackage{stfloats}
\usepackage{multicol}
\usepackage{empheq}
\usepackage{threeparttable}
\usepackage{color}
\usepackage{anyfontsize}
\usepackage{gensymb}
\usepackage{booktabs}
\usepackage{multirow}
\usepackage{array}
\usepackage{setspace}

\ifCLASSINFOpdf

\else

\fi

\begin{document}

\title{Design of A Single Antenna With Tunable In-Band RCS Null Through Load Impedance Control}

\author{Binchao~Zhang,
	Weidong~Hu,
	Fan~Yang,
	and Shenheng~Xu
\thanks{This work is supported in part by the National Natural Science Foundation of China under Grant 62401314, and in part by the YoungElite Scientists Sponsorship Program by CAST under Grant 2023QNRC001. (\textit{Corresponding author: Weidong Hu})}
\thanks{Binchao Zhang and Weidong Hu are with the school of Integrated Circuits and Electronics, Beijing Institute of Technology, Beijing 100081, and also are the Key Laboratory of Short-Range Radio Equipment Testing and Evaluation, Ministry of Industry and Information Technology and Terahertz Science Application Center (TSAC), Beijing Institute of Technology, Zhuhai, 519088, China. (Email: hoowind@bit.edu.cn, zhangbinchao@bit.edu.cn)}
\thanks{Fan Yang and Shenheng Xu are with Department of Electronic Engineering, Tsinghua University, Beijing, China.}
\vspace{-1cm}
}

\maketitle

\begin{abstract}
Reducing the in-band radar cross section (RCS) of antennas has been a widely concerned problem. However, most of works focus on RCS reduction of antenna arrays, or need to additionally increase the size of a single antenna. Therefore, this work presents a method to control the null of the in-band RCS by changing the load impedance of the antenna without additional aperture. Then, the monostatic and bistatic RCS can be effectively reduced. First, the relationship between the null position and the load impedance is calculated by analyzing the in-band scattering field of the antenna. Second, the performance of RCS null control is verified by the traditional patch antenna, whose operating frequency is designed at 2 GHz. The load impedance of the antenna is controlled by cascading an open-circuit stub of different lengths on the feeding line. Simulated results show that the proposed method can tune the null of in-band RCS from 0$^o$ to 60$^o$ under linear polarized normally incident plane wave. By setting the null reasonably, it can achieve 10 dB reduction for both monostatic and bistatic RCS at the same time. Moreover, the radiation performance of the antenna basically remains unchanged, and the gain is greater than 6.5 dBi. Measured results verify the effectiveness of the design method.
\end{abstract}

\begin{IEEEkeywords}
In-band, load impedance, RCS reduction, RCS null tunable, single antenna. 
\end{IEEEkeywords}

\IEEEpeerreviewmaketitle

\section{Introduction}
\indent With the rapid development of radar detection technology, the stealth performance of equipment platform has become an important issue for its survival capability. Particularly, the RCS caused by antenna scattering is the main contribution to the overall RCS of the equipment. Therefore, it is necessary to reduce the RCS of various antennas.\\
\indent In the past research, the RCS of antennas has been effectively reduced, whether it is in-band or out-of-band RCS. Common methods for out-of-band RCS reduction include the use of artificial magnetic conductors (AMC) \cite{SangDTAP2019}, broadband absorbers \cite{ZhangBTMTT2020,DhumalTEMC2023,MaTEMC2025}, and various multifunctional integrated metasurfaces \cite{ZhangBTAP2020,YuYAWPL2022,JinCTAP2021,ZhengQTAP2024,LiMAWPL2024}. Additionally, three common methods are used to reduce the in-band RCS of antennas. The earliest method involves optimizing the shape of the antenna, such as removing a portion of metal from antennas \cite{RajeshNTAP2017}, application of bionics \cite{JiangWAWPL2009}, and design of fractal patterns \cite{ThakareYBMAP2010}. Another popular method is to tune the scattering field of the antenna by using additional circuits at the feeding port \cite{ZhangZTAP2021,NakamotoNTAP2021,LiPTAP2022,ZhangZTAP2022,LiPTMTT2023,ZhaoCTAP2024,JiaYTAP2024}. The third method is to integrate antennas and metasurfaces, so that the radiation performance of the antenna and the RCS reduction ability of the metasurface are organically combined wihtout affecting each other, and then achieve both in-band radiation and stealth performance \cite{ZhangWAWPL2019,YangHTAP2021,GaoXAWPL2022,LvYTAP2022}. Moreover, both in-band and out-of-band RCS can be reduced by loading the metasurface onto the antenna \cite{JiKAWPL2024,WangPTAP2023,YangHTAP2023}.\\	
\indent However, the above mentioned works for reducing the in-band and out-of-band RCS require a certain aperture and a certain number of antenna elements, so that the RCS can be regulated by controlling the amplitudes and phases of each element's scattered electric field. It is rare to realize the RCS reduction for a single antenna, because it is hard to manipulate scattered electric field without array factor. An effective method is to arrange absorptive or diffusive metasurfaces around a single antenna to reduce the RCS of the antenna. Nonetheless, this approach requires artificially enlarging the aperture of the antenna to accommodate these metasurface elements \cite{ShiYTAP2019,ZhangQTAP2021,ZhangTAWPL2023}. Another approach is etching slots or adding branches in the antenna structure through optimization algorithms and characteristic mode theory \cite{WangWAWPL2010,ShiGTAP2023}. However, this method usually requires a complex optimization process, and the RCS null control is not realized.\\
\indent To fill this technological gap, this work presents a design method to control the null of the in-band RCS by controlling the load impedance of the antenna. First, the relationship between the null position and the load impedance is analyzed and calculated to give the load impedance value corresponding to differnet null position. Second, each required load impedance is obtained by optimizing the position and width of the cascaded open-circuit stub. Thus, the tunable in-band RCS null is realized without affect the antenna’s radiation performance and without increase the antenna's aperture size.
\vspace{-0.5cm}
\section{Theory and Design}
\vspace{-0.2cm}
\subsection{Theory}
\indent The scattering field $\vec{E}^{s}$ of the antenna typically comprises two parts, which are the structural mode scattering field $\vec{E}^{ss}$ and the antenna mode scattering field $\vec{E}^{as}$. These two scattering fields can be calculated using the scattering field of the antenna under the short and open loads, and they can be presented by
\begin{equation}\label{Formula-2}
\vec{E}^s(0)=\vec{E}^{ss}-\frac{1}{1+\Gamma_A}b_1^m\vec{E}_1^t
\end{equation}
\begin{equation}\label{Formula-3}
\vec{E}^s(\infty)=\vec{E}^{ss}+\frac{1}{1-\Gamma_A}b_1^m\vec{E}_1^t
\end{equation}
\indent Then, the antenna mode scattering field and the structural mode scattering field are
\begin{equation}\label{Formula-4}
\vec{E}^{as}=\frac{\Gamma_L}{1-\Gamma_A\Gamma_L}\frac{1-\Gamma_A^2}{2}\left[\vec{E}^s(\infty)-\vec{E}^s(0)\right]
\end{equation}
\begin{equation}\label{Formula-5}
\vec{E}^{ss}=\frac{(1-\Gamma_A)\vec{E}^s(\infty)+(1+\Gamma_A)\vec{E}^s(0)}{2}
\end{equation}
Based on the formulas (\ref{Formula-4}) and (\ref{Formula-5}), the scattering field of antennas $\vec{E}^s$ is simplified to
\begin{equation}\label{Formula-6}
\vec{E}^s=\frac{Z_L}{Z_A+Z_L}\vec{E}^s(\infty)+\frac{Z_A}{Z_A+Z_L}\vec{E}^s(0)
\end{equation}
Then, the amplitude and phase of the antenna’s scattering field can be controlled by varying the load impedance $Z_L$ for a fixed antenna structure. Because the $Z_A$, $\vec{E}^s(\infty)$, and $\vec{E}^s(0)$ are all constants. Concretely, if we want to obtian a RCS null at $\theta_{Null}$, that is $\vec{E}^s(Z_L)|_{\theta=\theta_{Null}}=0$, the terminated load impedance must satisfys
\begin{equation}\label{Formula-7}
Z_L|_{\theta=\theta_{Null}}=-Z_A*\frac{\vec{E}^s(0)_{\theta=\theta_{Null}}}{\vec{E}^s(\infty)_{\theta=\theta_{Null}}}.
\end{equation}
The following two conclusions are drawn from formula (\ref{Formula-7}):\\
(i) When $\theta_{Null}=0^o$, the minimum monostatic RCS is obtained.\\
(ii) By choosing different $\theta_{Null}$, the in-band RCS null can be controlled, which effectively reduce the RCS near the direction of $\theta_{Null}$.\\
However, the change of load impedance $Z_L$ will also inevitably lead to the deterioration of antenna impedance matching. Therefore, the variation range of $Z_L$ must be limited to meet the impedance matching performance, such as $|S_{11}|<-10$ dB.
\begin{figure}[t]
	\centering
	\subfigure []{
		\label{Element_PatchAnt_whole}
		\includegraphics[width=0.7\linewidth]{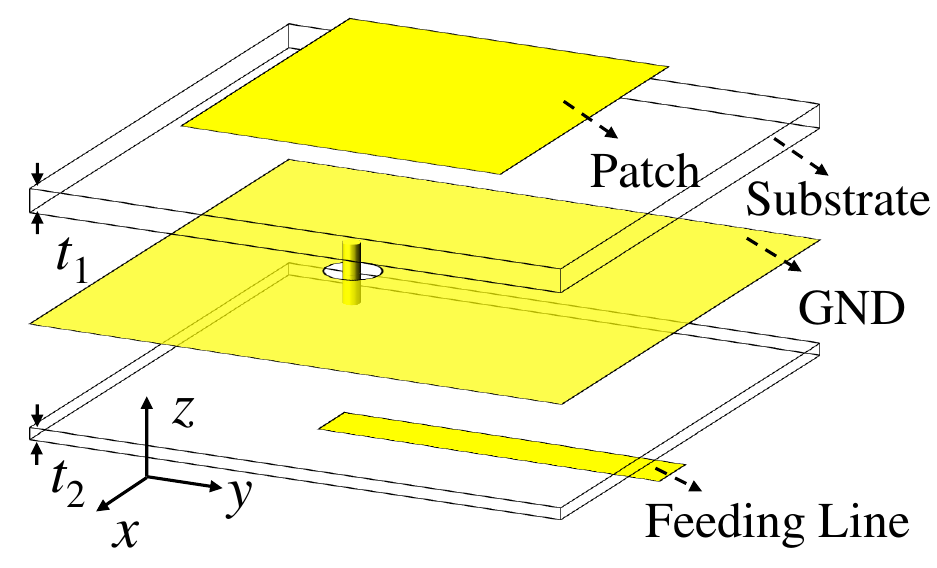}}
	\subfigure []{
		\label{Element_PatchAnt_bottom}
		\includegraphics[width=0.45\linewidth]{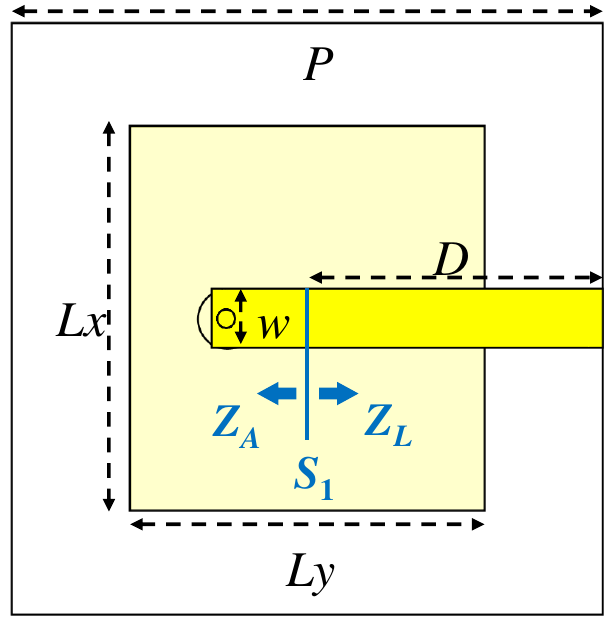}}
	\caption{Schematic diagram of the traditional patch antenna without controllable load impedance, (a) front view and (b) bottom view.}
	\label{Element_PatchAnt}
\end{figure}
\subsection{Antenna Design}
\indent A traditional patch antenna is used to verify the method of RCS null control by changing the load impedance, whose structure is shown in Fig. \ref{Element_PatchAnt}. The dielectric layers are Rogers RO3003 with a permittivity of 3, and their thicknesses are $t_1=2$ mm and $t_2=1$ mm, respectively. The antenna is designed to operate at a frequency of $f_0=2$ GHz, hence the dimensions of the patch are $L_x=45$ mm and $L_y=41.5$ mm, with a feed line width of $w=2.5$ mm and a via-hole radius of $r=0.5$ mm. The size of the antenna is $P=70$ mm, which is slightly small than half-wavelength.\\
\begin{figure}[t]
	\centering
	\includegraphics[width=0.85\linewidth]{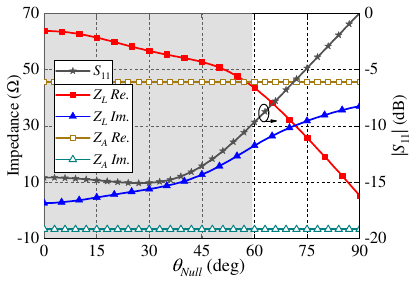}
	\caption{Calculated load impedance for different RCS nulls based on formula (\ref{Formula-6}), the simulated input impedance $Z_A$ and the calculated $S_{11}$.}
	\label{Z_Load_Required}
	\vspace{-0.5cm}
\end{figure}
Using the HFSS simulation software, the designed patch antenna is simulated with open boundary condition and a $y$-polarized normally incident plane wave. Then, the open-load scattered field $\vec{E}^s(\infty)$ and the short-load scattered field $\vec{E}^s(0)$ of the antenna are obtained. For the radiation simulation, a discrete port excitation is set at the reference plane $S_1$ (without loss of generality, the distance $D$ from the feed line terminal to $S_1$ is 23mm), thereby obtaining the antenna impedance $Z_A$=45.58-$j$6.69. Thus, according to formula (\ref{Formula-7}), the required load impedance to adjust the RCS null $\theta_{Null}$ from 0$^o$ to 90$^o$ at 2 GHz can be calculated, as shown in Fig. \ref{Z_Load_Required}. Moreover, the antenna input impedance $Z_A$ and the calculated $|S_{11}|=|((Z_L-Z_A ))/((Z_L+Z_A ))|$ are also presented in Fig. \ref{Z_Load_Required}. It is seen that under the restriction of $|S_{11}|<-10$ dB, the RCS null can be controlled in the range of $\pm60^o$, as shown in the gray area.\\
\begin{figure}[t]
	\centering
	\includegraphics[width=0.45\linewidth]{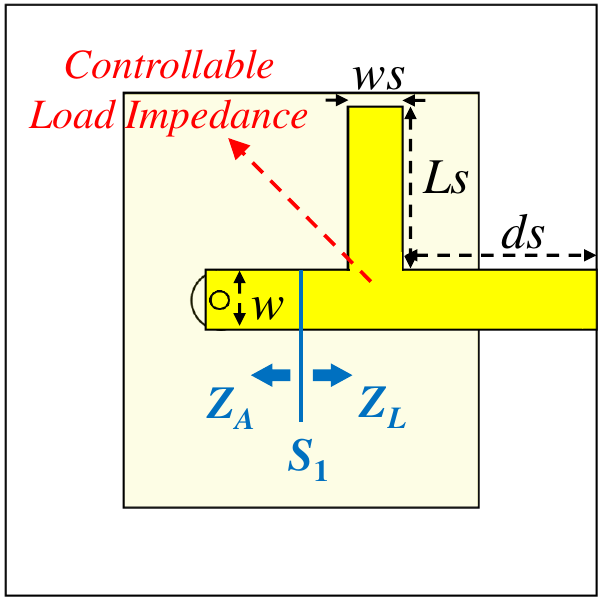}
	\caption{Designed antenna with controllable load impedance using open-circuit stub of different lengths.}
	\label{Element_Load_impedance}
	\vspace{-0.5cm}
\end{figure}
As shown in Fig. \ref{Element_Load_impedance}, the load impedance $Z_L$ is controlled by cascading open-circuit stub of different lengths $L_s$ at a distance $d_s$ from the terminal on the feeding line. Based on transmission line theory \cite{Pozar2005}, the load impedance $Z_L$ is
\begin{equation}\label{Formula-8}
Z_L=Z_{c}\frac{Z_L^{'}+jZ_{c}\tan\sqrt{\varepsilon_r}\beta(D-d_s)}{Z_{c}+jZ_L^{'}\tan\sqrt{\varepsilon_r}\beta(D-d_s)},
\end{equation}
\begin{equation}\label{Formula-9}
Z_L^{'}=Z_{c}\frac{Z_0+jZ_{c}\tan\sqrt{\varepsilon_r}\beta d_s}{Z_{c}+jZ_0\tan\sqrt{\varepsilon_r}\beta d_s}//-j\frac{Z_{cs}}{\tan\sqrt{\varepsilon_r}\beta L_s},
\end{equation}
where $Z_c$ and $Z_{cs}$ are the characteristic impedances of the feeding line and the cascaded open-circuit stub, respectively, which dependents on the line widths $w$ and $w_s$. The effective dielectric constant of the dielectric substrate is $\varepsilon_r$, and $\beta=2\pi⁄\lambda_0$ is the propagation constant, with $Z_0$ being 50 $\Omega$. Observing formulas (\ref{Formula-8}) and (\ref{Formula-9}), it is seen that the real part of the load impedance $Z_L$ is mainly determined by $Z_c$, while the imaginary part can be controlled by the parameter $d_s$.\\
\begin{figure}[t]
	\centering
	\subfigure []{
		\label{Z_Load_ds}
		\includegraphics[width=0.7\linewidth]{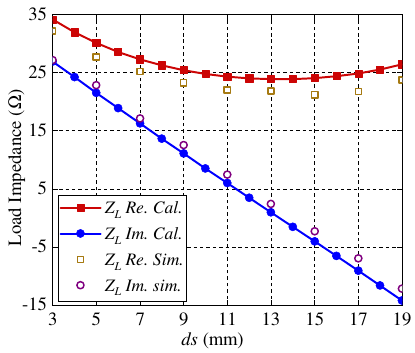}}
	\subfigure []{
		\label{Z_Load_w}
		\includegraphics[width=0.7\linewidth]{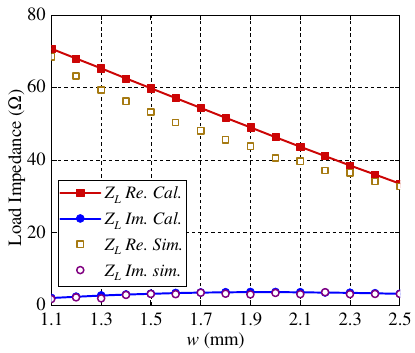}}
	\caption{Comparison of simulated and calculated results for load impedance with varying parameters: (a) different $d_s$, with $L_s=10$ mm, $w=w_s=2.5$ mm, (b) different $w$, with $L_s=5$ mm, $w_s=2.5$ mm, $d_s=10$ mm.}
	\label{Z_Load_paras}
\end{figure}
\begin{table}[t]
	\caption{Corresponding parameters for the required load impedance}
	\renewcommand\arraystretch{1.3}
	\begin{center} {\footnotesize
			\begin{tabular}{ccccccc}
				\hline
				$\theta_{Null}$ & Desired $Z_L$ ($\Omega$) & $w$ & $d_s$ & $L_s$  & $w_s$ & Sim. $Z_L$ ($\Omega$) \\
				\hline
				0$^o$ & 63.8+$j$2.6 & 1.4 & 9.5 & 4 & 2.5 & 63.9+$j$2.8 \\
				\hline
				15$^o$ & 61.4+$j$4.6 & 1.45 & 8.8 & 4 & 2.5 & 62.7+$j$4.9 \\
				\hline
				30$^o$ & 56.7+$j$7.6 & 1.65 & 7.5 & 4 & 2.5 & 57.2+$j$7.6 \\
				\hline
				45$^o$ & 52.7+$j$12.7 & 1.9 & 4 & 4 & 2.5 & 53.9+$j$12.4 \\
				\hline
				60$^o$ & 43.5+$j$23.1 & 1.1 & 8.3 & 10 & 2.5 & 42.6+$j$23.2 \\
				\hline
		\end{tabular} }
	\end{center}
	\label{table1}
\end{table}
\begin{figure}[t]
	\centering
	\subfigure []{
		\label{RCS_Null_Cal_Sim}
		\includegraphics[width=0.7\linewidth]{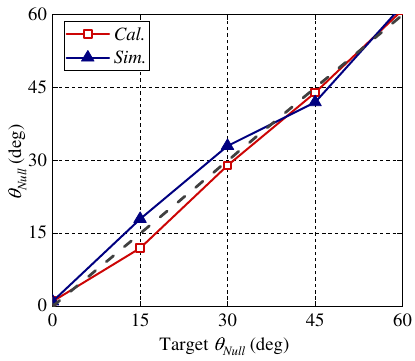}}
	\subfigure []{
		\label{S11_Gain_Sim}
		\includegraphics[width=0.7\linewidth]{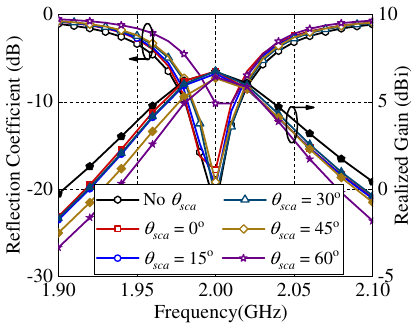}}
	\caption{(a) Comparison of simulated and calculated RCS nulls, (b) simulated reflection coefficients and gains of RCS nulls at $\theta_{Null}=(0^o,15^o,30^o,45^o,60^o)$.}
	\label{RCS_S11_Gain_Sim}
\end{figure}
As shown in Fig. \ref{Z_Load_ds}, the distance $d_s$ of the cascaded open-circuit stub from the terminal has a significant impact on the imaginary part of the load impedance and a slightly impact on the real part. Increasing $d_s$ can effectively decrease the imaginary part of the load impedance. Fig. \ref{Z_Load_w} shows the relationship between the feed line width $w$ and the load impedance. It is observed that increasing $w$ can effectively reduce the real part of the load impedance, while the imaginary part remains relatively unchanged. Therefore, for the required load impedance $Z_L|_{\theta=\theta_{Null}}$, one can first adjust the imaginary part of the impedance by changing $d_s$, and then fit the real part by changing $w$ to obtain the desired load impedance.
\vspace{-0.2cm}
\subsection{RCS Null Regulation}
\begin{figure}[t]
	\centering	
	\includegraphics[width=0.75\linewidth]{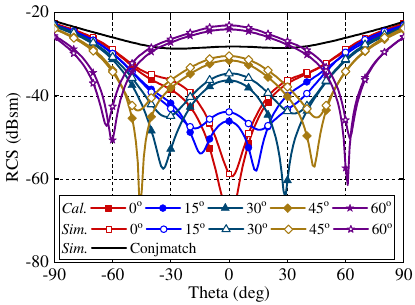}
	\caption{Calculated and simulated RCS patterns at 2 GHz for different nulls at $\theta_{Null}=(0^o,15^o,30^o,45^o,60^o)$ and also the simulated RCS pattern of the antenna with conjugate matched load.}
	\label{RCS_Pattern_Sim}
\end{figure}

\begin{figure}[!t]
	\centering
	\subfigure []{
		\label{3D_RCS_pattern_withoutLoad}
		\includegraphics[width=0.4\linewidth]{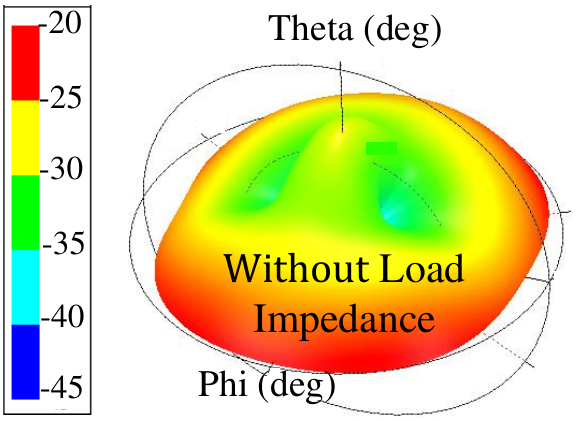}}
	\subfigure []{
		\label{3D_RCS_pattern_theta30}
		\includegraphics[width=0.4\linewidth]{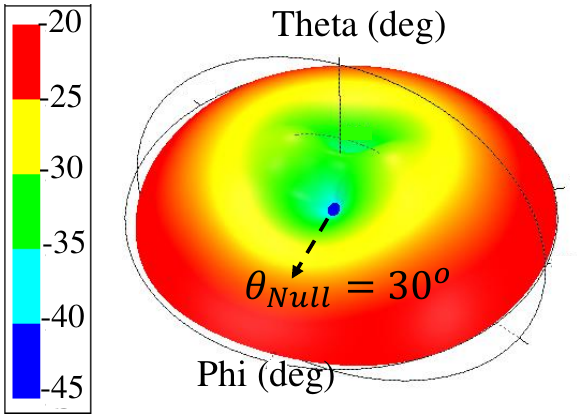}}
	\subfigure []{
		\label{3D_RCS_pattern_theta60}
		\includegraphics[width=0.4\linewidth]{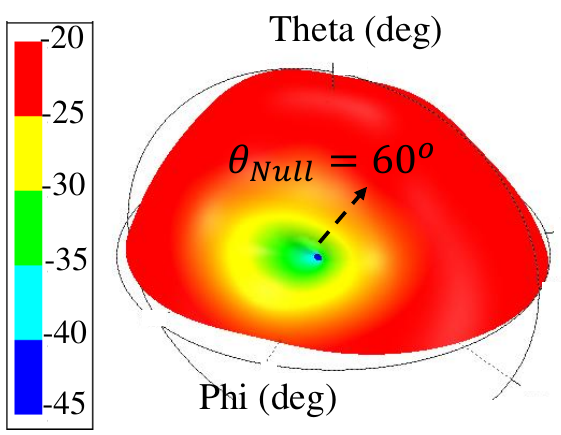}}
	\caption{Simulated 3D RCS patterns of the designed antenna with and without the impedance, (a) without load impedance, (b) with load impedance for $\theta_{Null}=30^o$, and (c) with load impedance for $\theta_{Null}=60^o$.}
	\label{3D_RCS_patterns}
\end{figure}

\indent Following the aforementioned method, by adjusting $d_s$ and $w$, parameter combinations corresponding to the RCS nulls at $\theta_{Null}=(0^o,15^o,30^o,45^o,60^o)$ are obtained, as shown in Table \ref{table1}. It is seen that the simulated load impedance has little deviation from the desired values. Using these parameter combinations, the simulated RCS nulls are obtained and compared with the theoretical calculated values, as shown in Fig. \ref{RCS_Null_Cal_Sim}, and the results agree well. Here, the calculated results involve substituting the simulated $Z_L$ from Table \ref{table1} into formula (\ref{Formula-6}) to calculate the scattering electric field, and then calculating the RCS using the formula of $\sigma=4\pi R^2|\vec{E}^s|^2/|\vec{E}^i|^2$ \cite{Knott2004}, where the distance $R=1$ m and the incident electric field strength $|\vec{E}^i|=1$ V/m. In addition, Fig. \ref{S11_Gain_Sim} shows the reflection coefficients and realized gains of the antenna when tunning the RCS nulls. It is seen that the antenna's reflection coefficients remain below -10 dB within 1.98$\sim$2.02 GHz, and the realized gain is stable. Additionally, when the RCS null is at 60$^o$, $S_{11}$ approaches -10 dB, suggesting that the upper limit for null tunning is around 60$^o$.\\
\indent The calculated and simulated bistatic RCS patterns at 2 GHz when the nulls at $\theta_{Null}=(0^o,15^o,30^o,45^o,60^o)$ are presented in Fig. \ref{RCS_Pattern_Sim}, which are in good agreement. From these results, it is concluded that the null of the in-band RCS can be flexibly controlled by changing the load impedance. Moreover, compared to the conjugate matched load, the monostatic RCS is significantly reduced when the null changes from 0$^o$ to 45$^o$. In particular, when null is set to 0$^o$, the monostatic RCS is more than 30 dB lower than conjugated matched load. Therefore, by adjusting the load impedance, the monostatic and bistatic RCS can be reduced simultaneously. Moreover, the simulated 3D RCS patterns are presented in Fig. \ref{3D_RCS_patterns}. It is intuitively seen that the RCS null can be controlled by adding and adjusting different load impedances.
\begin{figure}[t]
	\centering
	\subfigure []{
		\label{Prototype_top}
		\includegraphics[width=0.98\linewidth]{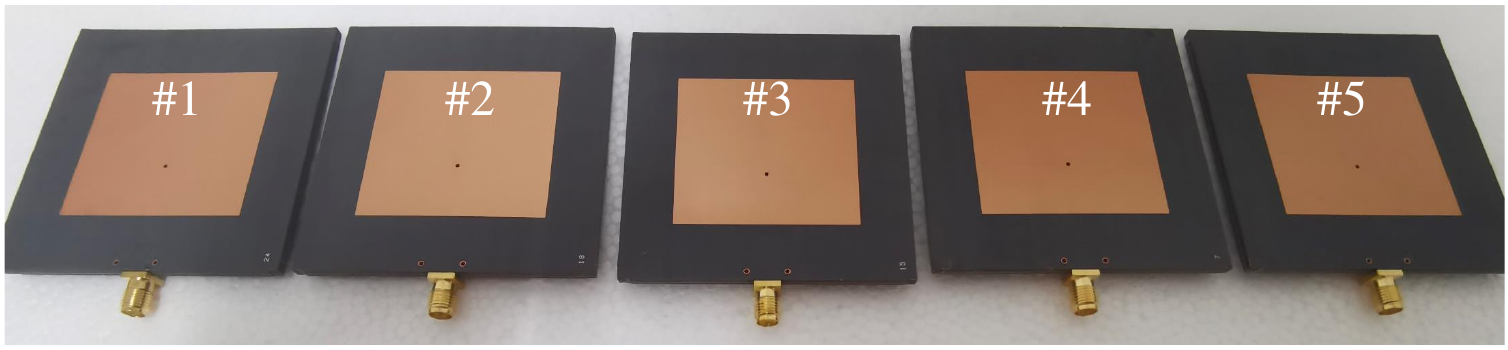}}
	\subfigure []{
		\label{Prototype_bottom}
   		\includegraphics[width=0.98\linewidth]{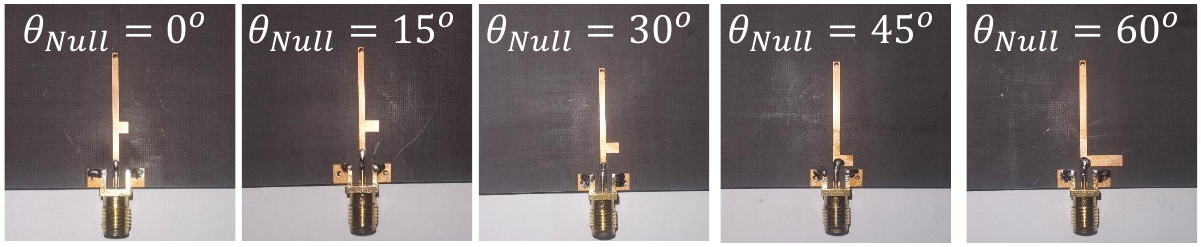}}
	\caption{Fabrication prototypes, (a) top view and (b) bottom view.}
	\label{Prototype}
\end{figure}
%\begin{figure}[t]
%	\centering
%	\label{Prototype_bottom}
%   	\includegraphics[width=0.9\linewidth]{figure/Prototype_bottom.pdf}
%	\caption{Fabrication prototypes for different RCS nulls.}
%	\label{Prototype}
%\end{figure}
\begin{table}[t]
	\caption{Optimized parameters for the different RCS nulls}
	\renewcommand\arraystretch{1.3}
	\begin{center} {\footnotesize
			\begin{tabular}{cccccc}
				\hline
				Ant. Order & ~$\theta_{Null}$~ & ~~$w$~~ & ~~$d_s$~~ & ~~$L_s$~~ & ~~$w_s$~~ \\
				\hline
				\#1 & 0$^o$ & 1.4 & 12.8 & 2 & 2.5 \\
				\hline
				\#2 & 15$^o$ & 1.3 & 12 & 3 & 2.5 \\
				\hline
				\#3 & 30$^o$ & 1.4 & 10 & 3 & 2.5 \\
				\hline
				\#4 & 45$^o$ & 1.6 & 6 & 3 & 2.5 \\
				\hline
				\#5 & 60$^o$ & 1.5 & 6 & 8 & 2.5 \\
				\hline
		\end{tabular} }
	\end{center}
	\label{table2}
\end{table}
\begin{figure}[!t]
	\centering
	\subfigure []{
		\label{MeasureSetup-Radiation}
		\includegraphics[width=0.48\linewidth]{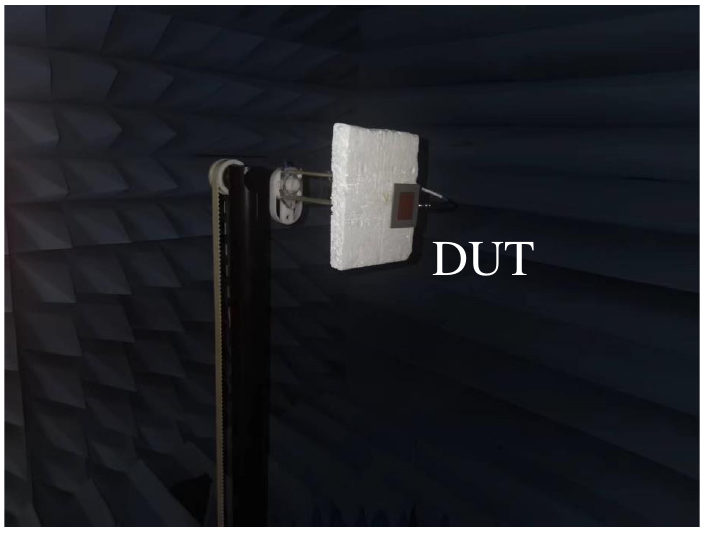}}
	\subfigure []{
		\label{MeasureSetup-RCS}
		\includegraphics[width=0.48\linewidth]{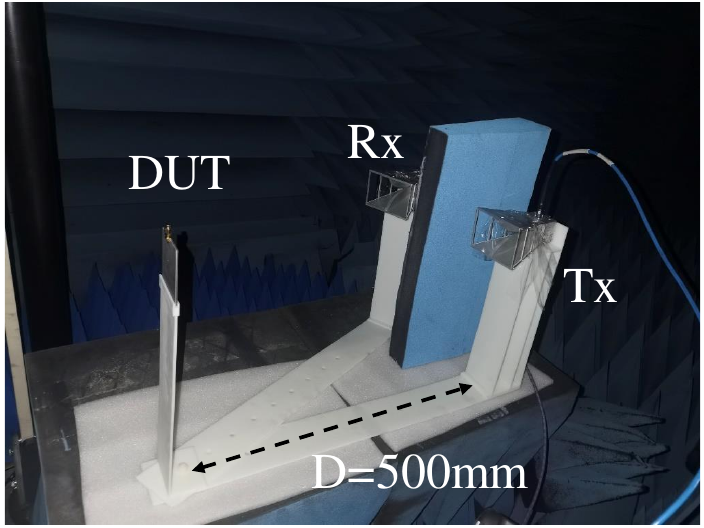}}
	\caption{Measurement setup for (a) radiation and (b) scattering performance.}
	\label{MeasureSetup}
\end{figure}
\section{Fabrication and Measurement}
\indent Five patch antennas for the nulls at $\theta_{Null}=(0^o,15^o,30^o,45^o,60^o)$ are fabricated to verify the proposed design method, as shown in Fig. \ref{Prototype}. These five prototypes have the same radiation patch but the different loads. It should be noted that compared to the simulation structure, the fabricated prototypes need to introduce the SMA and the patch for grounding. Thus, the parameters in Table \ref{table1} need to be re-optimized, which are listed in Table \ref{table2}. Then, the fabricated prototypes are placed in the anechoic chamber to measure the radiation and scattering performance, as shown in Fig. \ref{MeasureSetup}. The designed antennas operate as a transmitting antenna when the radiation performance is measured, as shown in Fig. \ref{MeasureSetup-Radiation}. When the scattering performance is measured, a transmitting linear polarization antenna Tx serves to normally irradiate the designed antenna, while the receiving linear polarization antenna Rx, rotates around to measure the scattering pattern, as shown in Fig. \ref{MeasureSetup-RCS}. The distance $D$ between the designed antenna and the Tx antenna is 500 mm to guarantee far-field conditions. Then, the measured reflection coefficients at different scattering directions are used to build the scattering pattern of the antenna under test, which can be utilized as the evaluation basis of RCS reduction performance \cite{ZhangZTAP2023}. \\
\begin{figure}[!t]
	\centering
	\subfigure []{
		\label{S11_Gain_Meas}
		\includegraphics[width=0.7\linewidth]{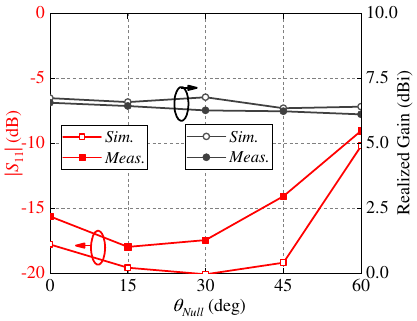}}
	\subfigure []{
		\label{Gain_Pattern_Meas}
		\includegraphics[width=0.7\linewidth]{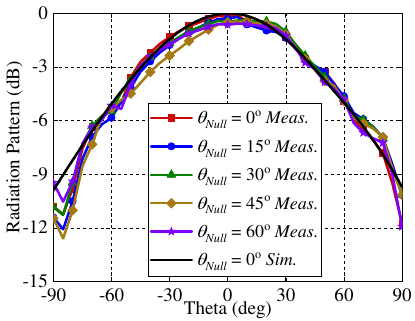}}
	\caption{Simulated and measured radiation performance for RCS nulls at $\theta_{Null}=(0^o,15^o,30^o,45^o,60^o)$, (a) reflection coefficients and realized gains, (b) normalized radiation patterns.}
	\label{MeasRadiation}
\end{figure}
\indent The measured radiation results are presented in Fig. \ref{S11_Gain_Meas}. Compared with the simulated results, the measured reflection coefficients and realized gains are agreed well. The slight difference is due to the errors caused by machining and welding. Moreover, the measured normalized radiation patterns under different RCS nulls are shown in Fig. \ref{Gain_Pattern_Meas}, which are basically stable when the null of RCS changes.\\
\begin{figure}[!t]
	\centering
	\includegraphics[width=0.7\linewidth]{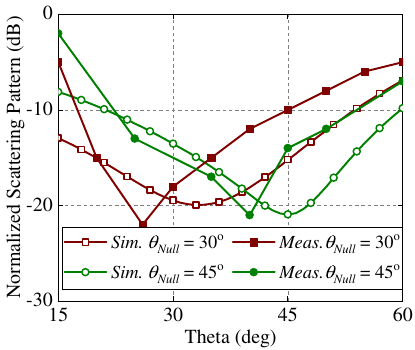}
	\caption{Simulated and measured normalized scattering patterns for $\theta_{Null}=(30^o,45^o)$.}
	\label{RCS_Meas}
\end{figure}
\indent When assessing the scattering performance, it should be mentioned that due to the limitations of the arch frame, the range of the scattering pattern is confined within 15$^o$ to 60$^o$, and thus the RCS nulls at 30$^o$ and 45$^o$ are measured, as presented in Fig. \ref{RCS_Meas}. The measured results show that the designed antennas can control the in-band and co-polarization RCS null, although the measured nulls are slightly shifted compared to the simulated results. The reason is that the welding of SMA will inevitably introduces additional parasitic parameters, which will enlarge the load impedance, so that the nulls will shift to a lower angle.\\
\begin{table}[!t]
	\caption{Performance Comparison}
	\renewcommand\arraystretch{1.1}
	\begin{center} {\footnotesize
			\begin{tabular}{cccccc}
				\hline
				Ref. & \begin{tabular}[c]{@{}c@{}}Size\\($\lambda_0$) \end{tabular} & \begin{tabular}[c]{@{}c@{}}Gain\\(dBi) \end{tabular} & \begin{tabular}[c]{@{}c@{}}RCSR\\Method \end{tabular} & \begin{tabular}[c]{@{}c@{}}RCSR\\Band \end{tabular} & \begin{tabular}[c]{@{}c@{}}RCS Null\\Tuning\end{tabular} \\
				\hline
				\cite{ShiYTAP2019} & 1.05$\times$1.05 & 6.4 & \begin{tabular}[c]{@{}c@{}}Metasurfaces\\Surrounded \end{tabular}  & In/Out & No \\
				\hline
				\cite{ZhangQTAP2021} & 1.57$\times$1.57 & 7.66  & \begin{tabular}[c]{@{}c@{}}Metasurfaces\\Surrounded \end{tabular} & In/Out  & No  \\
				\hline
				\cite{ZhangTAWPL2023}  & 2.48$\times$2.48 & 7 & \begin{tabular}[c]{@{}c@{}}Metasurfaces\\Surrounded \end{tabular} & In/Out  &  No\\
				\hline
				\cite{WangWAWPL2010}  & 0.8$\times$0.64 & 9.37 & Grooves & Out  &  No\\
				\hline
				\cite{ShiGTAP2023}  & 0.63$\times$0.63 & 6.9 & \begin{tabular}[c]{@{}c@{}}Grooves and\\Stubs \end{tabular} & In  &  No\\
				\hline
				\begin{tabular}[c]{@{}c@{}}This\\Work \end{tabular} & 0.47$\times$0.47 & 6.5  & \begin{tabular}[c]{@{}c@{}}Changing Load\\Impedance \end{tabular}  & In &  Yes\\
				\hline
		\end{tabular} }
	\end{center}
	\label{Performance_Comparison}
\end{table}
\indent Recent work on RCS reduction for single antennas are compared in Table \ref{Performance_Comparison}. One class is to manipulate incident waves by placing metasurfaces around the antenna, thus enabling both in-band and out-of-band RCS reduction. However, it is obvious that the antenna size needs to be increased to accommodate a certain number of metasurface elements. On the other hand, by means of characteristic mode theory, the in-band or out-of-band RCS can be effectively reduced by etching grooves and adding stubs on the radiation structure or the ground, and it also avoids expanding the antenna size. Unfortunately, none of the above-mentioned methods can control the RCS null. Therefore, the innovative contribution of this work lies in the fact that a single antenna can realize RCS null controlling without the necessary of forming an array or additional apeature. Then, the monostatic and bistatic RCS can be effectively reduced by changing the load impedance of the antenna.
\vspace{-1em}
\section{Conclusion}
A design method to control the RCS null by controlling the load impedance is proposed in this work. Through analyzing and calculating the relationship between the null and the load impedance, each required RCS null can be realized by optimizing the position and width of the cascaded open-circuit stub. Thus, the tunable in-band RCS null is realized without affect the antenna’s radiation performance. This design method can provide an effective solution for the reduction of the monostatic and bistatic RCS of a single antenna.

\ifCLASSOPTIONcaptionsoff
  \newpage
\fi

\end{document}